\documentclass[aps,prd,twocolumn,tightenlines,superscriptaddress,nofootinbib,preprintnumbers]{revtex4-2}

\usepackage{amsmath}
\usepackage{amssymb}
\usepackage{latexsym}
\usepackage{graphicx}
\usepackage{slashed}
\usepackage{pifont}
\usepackage{color}
\usepackage{multirow}
\usepackage[usenames,dvipsnames]{xcolor}
\usepackage[caption=false]{subfig}
\usepackage[colorlinks=True, citecolor=blue, urlcolor=blue, linkcolor=blue]{hyperref}
\usepackage{cleveref}
\usepackage{orcidlink}
\usepackage{float}
\usepackage{booktabs}
\usepackage[normalem]{ulem}

\hypersetup{colorlinks=true,citecolor=[rgb]{0,0,0.8},
linkcolor=[rgb]{0,0,0.8},urlcolor=[rgb]{0,0,0.8}}

\begin{document}

\title{Magic Production and the Fundamental Forces of Nature}

\author{Jiunn-Wei Chen}
\email{jwc@phys.ntu.edu.tw}
\affiliation{Department of Physics and Center for Theoretical Physics,
National Taiwan University, Taipei, Taiwan 106}
\affiliation{Physics Division, National Center for Theoretical Sciences, Taipei 10617, Taiwan}
\affiliation{InQubator for Quantum Simulation (IQuS), Department of Physics, University of Washington, Seattle, WA 98195, USA.}

\author{Yu-Hao Li}
\email{r13222023@g.ntu.edu.tw}
\affiliation{Department of Physics and Center for Theoretical Physics,
National Taiwan University, Taipei, Taiwan 106}

\preprint{IQuS@UW-21-140}

\date{\today}

\begin{abstract}

Quantum magic is a resource for universal quantum computation and may
also provide a probe of fundamental interactions. We study its
production in forward scattering of two distinct Standard Model
fermions in the chirality basis. In the single-mediator-exchange
approximation, magic production vanishes for gluon, photon, $W$-boson,
Higgs-boson, and graviton exchange. For $Z$-boson exchange in purely
leptonic scattering, it also vanishes at $\sin^2\theta_W=1/4$, close
to the phenomenological value. By contrast, $Z$-boson exchange
involving elementary quarks generally produces nonzero magic. These
patterns raise the question of whether the construction of the
fundamental forces encodes, at least in some channels, a tendency to
suppress magic production.

\end{abstract}

\maketitle

\section{Introduction}
\label{sec:intro}
Entanglement is often used to characterize the complexity of a quantum
system. However, even highly entangled states can be simulated
efficiently on a classical computer if they are generated by circuits
involving only stabilizer-state preparations, Clifford gates, and Pauli
measurements. The required computational resources then scale
polynomially with the system size. This result is known as the
Gottesman--Knill theorem~\cite{Gottesman:1998hu,Aaronson:2004xuh}.

Universal quantum computation, however, requires operations beyond the
Clifford group. Nonstabilizerness, or quantum magic, quantifies a
different aspect of quantum complexity from entanglement and is a
resource for universal quantum computation.

High entanglement and high magic may make classical simulation more
difficult, but neither guarantees a quantum advantage. For example,
entanglement is not necessary for some restricted oracle tasks, while
circuits with extensive entanglement and magic can still be simulated
efficiently for particular observables~\cite{Biham:2003,Aaronson:2004xuh,Fux:2025}.

Recent studies of entanglement and magic in nuclear and particle physics
are motivated not only by their relevance to computational resources,
but also by the possibility that they may reveal properties of
fundamental interactions. In nuclear physics, Robin and Savage studied
the ``magic power'' of the scattering matrix and found distinctive
patterns in nucleon--nucleon and hyperon--nucleon scattering using
empirical phase shifts. They also identified the $\Sigma^-$ as a
possible catalyst for enhanced magic production and entanglement
spreading in dense matter~\cite{Robin:2024oqc}. Complementary shell-model
calculations found substantial proton--neutron entanglement and magic
in $sd$-shell nuclei, including nuclei near shape coexistence and level
inversion~\cite{Brokemeier:2024lhq}. In particle physics, QED has been
found to produce little magic~\cite{Liu:2025qfl}. In electroweak
theory, both angle-averaged magic production~\cite{Liu:2025bgw} and the
variation of R\'enyi mutual information~\cite{Cao:2026aye} in
charged-lepton scattering are minimized near the measured weak mixing
angle. Magic production in gluon--gluon and graviton--graviton scattering
and their supersymmetric extensions has also been studied
~\cite{Gargalionis:2025iqs}.

Here we investigate whether magic production can reveal features of the
fundamental interactions: the strong, weak, electromagnetic, and
gravitational forces. We study high-energy scattering of Standard Model
fermions in the forward limit, where $t$-channel contributions are
kinematically enhanced. We find that, to good accuracy, magic
production vanishes in nearly all the processes considered. A notable
exception is $Z$-boson exchange involving elementary quarks, which
produces nonzero magic. This pattern raises the question of whether
magic production reflects a deeper design encoded in the fundamental
forces.

\section{Vanishing Magic Production in Forward Scattering}
\label{sec:H+QC}

An $n$-qubit pure stabilizer state is the unique common $+1$ eigenstate
of $n$ independent, mutually commuting, signed Hermitian Pauli
generators. Their products form an Abelian stabilizer group of $2^n$
elements that does not contain $-I$. The Clifford group consists of
unitary operators that map Pauli operators to Pauli operators under
conjugation. Up to overall phases, it is generated by the one-qubit
Hadamard ($H$) and phase ($S$) gates and the two-qubit controlled-NOT
(CNOT) gate, whose matrix representations are
\begin{align*}
H &=
\frac{1}{\sqrt{2}}
\begin{pmatrix}
1 & 1\\
1 & -1
\end{pmatrix},
&
S &=
\begin{pmatrix}
1 & 0\\
0 & i
\end{pmatrix},
&
\mathrm{CNOT} &=
\begin{pmatrix}
1 & 0 & 0 & 0\\
0 & 1 & 0 & 0\\
0 & 0 & 0 & 1\\
0 & 0 & 1 & 0
\end{pmatrix}.
\end{align*}

The six one-qubit stabilizer states are the eigenstates of the Pauli
operators $X$, $Y$, and $Z$. There are 60 two-qubit stabilizer states:
36 product states formed from pairs of one-qubit stabilizer states and
24 maximally entangled states.

We consider forward scattering, $AB\to AB$, between two distinct
fermions, with their spins expressed in the right- and left-handed
chirality basis
$\{|R_A\rangle,|L_A\rangle\}\otimes\{|R_B\rangle,|L_B\rangle\}$.
For an initial stabilizer state $|i\rangle$, the transition matrix
$\mathcal{T}$ produces the normalized final spin state
\begin{equation}
|f\rangle =
\frac{\mathcal{T}|i\rangle}
{\sqrt{\langle i|\mathcal{T}^{\dagger}\mathcal{T}|i\rangle}} \, .
\end{equation}

Although several measures of magic, or nonstabilizerness, are
available, we use the second stabilizer R\'enyi entropy
~\cite{Leone:2021rzd} as a diagnostic for pure states:
\begin{equation}
 M_2(|f\rangle)=-\log\left[
 \frac{1}{2^n}
 \sum_{P\in\{I,X,Y,Z\}^{\otimes n}}
 |\langle f|P|f\rangle|^4
 \right].
 \label{eq:M2}
\end{equation}
For normalized pure states, $M_2\geq 0$, with equality precisely for
stabilizer states. Different diagnostics may assign different values
to nonstabilizer states, but our conclusion of vanishing magic is
independent of the chosen pure-state measure, provided that it
vanishes precisely for stabilizer states.

The spin-zero Higgs couples to fermions through the Yukawa operator
\[
\bar{\psi}\psi
=
\bar{\psi}_L\psi_R+\bar{\psi}_R\psi_L,
\]
which flips chirality at each fermion line. Thus, in the
forward-scattering limit, the two-fermion transition operator is
\begin{equation}
\label{Th}
\mathcal{T}_h \propto X\otimes X .
\end{equation}
Since $X=HSSH$, this operator is Clifford and maps stabilizer states
to stabilizer states. Therefore, a single Higgs exchange produces no
magic in the forward-scattering limit.

For exchange of a spin-one gauge boson with chiral current
\[
g_L^i\,\bar{\psi}_{L}^{\,i}\gamma^\mu\psi_{L}^{\,i}
+
g_R^i\,\bar{\psi}_{R}^{\,i}\gamma^\mu\psi_{R}^{\,i},
\]
the forward-scattering transition operator takes the form
\begin{equation}
\label{T}
\mathcal{T}
\propto
\operatorname{diag}(g_R^A,g_L^A)
\otimes
\operatorname{diag}(g_R^B,g_L^B).
\end{equation}
For real chiral couplings, zero magic production in the forward limit
occurs for four coupling structures:
\begin{align*}
\mathbf{V}:&\quad g_R=g_L
&&\text{(vector)},\\
\mathbf{A}:&\quad g_R=-g_L
&&\text{(axial vector)},\\
\mathbf{R}:&\quad g_L=0
&&\text{(right-handed)},\\
\mathbf{L}:&\quad g_R=0
&&\text{(left-handed)}.
\end{align*}
The corresponding single-fermion factors are
\begin{equation}
\label{diag}
\operatorname{diag}(g_R^i,g_L^i)
\propto
I,\quad
Z,\quad
\frac{I+Z}{2},\quad
\frac{I-Z}{2},
\end{equation}
for $\mathbf{V}$, $\mathbf{A}$, $\mathbf{R}$, and $\mathbf{L}$ couplings,
respectively. The couplings may differ for the two fermions,
$i=A,B$.

Using Eq.~(\ref{diag}), the transition matrix in Eq.~(\ref{T}) is a
tensor product of single-qubit Clifford operators and Pauli
projectors. In particular, $Z=S^2$ is Clifford, while $(I\pm Z)/2$
project onto the $Z$ eigenstates. Clifford operators map stabilizer
states to stabilizer states, and Pauli projections also map them to
stabilizer states for any nonzero outcome. Thus, these transition
operators do not generate magic from stabilizer inputs.

Within the Standard Model gauge group
$SU(3)_c\times SU(2)_L\times U(1)_Y$, the $SU(3)_c$ and $SU(2)_L$
interactions satisfy the zero-magic criterion identified above: the
strong interaction is vectorlike, whereas $SU(2)_L$ couples only to
left-handed fermions. This observation motivates a
quantum-information perspective on the chiral structure of the weak
interaction, although it does not by itself explain why $SU(2)_L$ is
left-handed.

By contrast, the hypercharge interaction $U(1)_Y$ does not satisfy this
criterion. For the electron, its chiral couplings obey
$g_L=g_R/2$, which belongs to none of the
$\mathbf{V}$, $\mathbf{A}$, $\mathbf{R}$, or $\mathbf{L}$ classes.

This motivates us to ask whether electroweak symmetry breaking
reorganizes the gauge couplings into forms that satisfy the
zero-magic criterion. When the Higgs field acquires a vacuum
expectation value, the neutral gauge fields of $SU(2)_L$ and $U(1)_Y$
mix to form the photon and the $Z$ boson. The $W^\pm$ bosons retain
purely left-handed couplings, while the photon couples vectorially;
both therefore belong to zero-magic classes. The $Z$-boson couplings,
\begin{equation}
g_L^f = T_3^f-Q_f\sin^2\theta_W,
\qquad
g_R^f = -Q_f\sin^2\theta_W,
\end{equation}
are generally chiral mixtures; the common factor $g/\cos\theta_W$ is
omitted. For neutrinos, $Q_\nu=0$ and $g_R^\nu=0$, so the coupling is
purely left-handed. For electrons, $(T_3,Q)=(-1/2,-1)$, and the
$Z$ coupling is purely left-handed, axial, or right-handed at
$\sin^2\theta_W=0$, $1/4$, or $1/2$, respectively. Notably,
$\sin^2\theta_W=1/4$ is close to its phenomenological
$\overline{\mathrm{MS}}$ values: approximately $0.24$ below
$1~\mathrm{GeV}$ and $0.23$ at the $Z$ pole
~\cite{ParticleDataGroup:2026mpi}.

In Møller scattering, photon and $Z$ exchange contribute coherently.
Although either exchange separately can produce zero magic in the
forward limit for special couplings, their sum does not generally
retain this property. Ref.~\cite{Liu:2025bgw} used the magic
production averaged over scattering angle and initial stabilizer
states to identify the value of $\sin^2\theta_W$ that minimizes it,
finding agreement with the phenomenological value over a broad energy
range. Subsequent analyses showed that the inferred minimum depends
on the chosen basis~\cite{Li:2026zum} and magic measure
~\cite{Shen:2026lsn}.

At $\sin^2\theta_W=1/4$, the forward $Z$ couplings to elementary
quarks satisfy
\[
\frac{g_L^u}{g_R^u}=-2,
\qquad
\frac{g_L^d}{g_R^d}=-5.
\]
Neither ratio corresponds to the $\mathbf{V}$, $\mathbf{A}$,
$\mathbf{R}$, or $\mathbf{L}$ classes, so the zero-magic criterion
fails for elementary-quark external states.

For nucleon external states, however, the effective couplings are
different. At zero momentum transfer, the proton's vector coupling is
proportional to $1/2-2\sin^2\theta_W$, which vanishes at
$\sin^2\theta_W=1/4$. Its remaining coupling is therefore purely
axial, corresponding to the $\mathbf{A}$ class. Other form-factor
terms, proportional to $\sigma^{\mu\nu}q_\nu$,
$\sigma^{\mu\nu}q_\nu\gamma^5$, or $q^\mu\gamma^5$, vanish as
$q^\mu\to0$.

The neutron current is proportional to
$\bar n\gamma^\mu(1-g_A\gamma^5)n$, where $g_A\simeq1.27$. In the
chiral basis,
\[
\bar n\gamma^\mu(1-g_A\gamma^5)n
=(1+g_A)\bar n_L\gamma^\mu n_L
 +(1-g_A)\bar n_R\gamma^\mu n_R .
\]
Thus, the neutron coupling is predominantly left-handed, with a
right-handed amplitude about $12\%$ of the left-handed one.

The zero-magic condition is generally violated for baryons containing
valence strange quarks. Protons and neutrons may nevertheless be of
particular interest because of their abundance in nature. We therefore
consider low-energy nuclear effective field theory (EFT)
~\cite{Gasser:1983yg,Bernard:1995dp,Weinberg:1990rz,Weinberg:1991um,Weinberg:1992yk,Kaplan:1998tg,Kaplan:1998we,Beane:2000fx,Bedaque:2002mn,Meissner:2014lgi,Hammer:2012id},
in which heavier baryons are integrated out and nucleons are the
relevant degrees of freedom. Since the nucleon mass couples
left- and right-handed components, spin is a more natural basis than
chirality.

The leading spin-independent nucleon contact interaction,
$(N^\dagger N)^2$, is invariant under Wigner's spin-isospin $SU(4)$
symmetry~\cite{Mehen:1999qs,Beane:2018oxh}. In the two-nucleon spin
space, its transition operator is proportional to the identity,
\begin{equation}
\label{eq:Tstrong}
\mathcal{T}_{\rm strong} \propto I\otimes I ,
\end{equation}
so this term does not generate magic~\cite{Low:2026evp}. The
spin-dependent contact interaction,
$(N^\dagger\boldsymbol{\sigma}N)^2$, is suppressed relative to the
spin-independent term by a factor of order $1/N_c^2$ in the
large-$N_c$ expansion, where $N_c$ is the number of colors
~\cite{Kaplan:1995yg}.

By contrast, $Z$-boson exchange between two protons with an
axial-vector ($\mathbf{A}$-class) coupling produces a spin-dependent
contribution, $(N^\dagger\boldsymbol{\sigma}N)^2$. In the spin basis,
its leading operator is proportional to
\begin{equation}
\label{eq:TZ}
\mathcal{T}_{Z} \propto
X\otimes X+Y\otimes Y+Z\otimes Z .
\end{equation}
This operator can generate nonzero magic from suitable stabilizer
inputs. Thus, the zero-magic condition is violated for this
$Z$-exchange contribution.

We now turn to gravity. A spin-2 graviton couples universally to the
fermion energy-momentum tensor,
\begin{equation}
  T_{\mu\nu}
  = \frac{i}{4}\,
  \bar{\psi}\left(
  \gamma_\mu \overleftrightarrow{D}_\nu
  +\gamma_\nu \overleftrightarrow{D}_\mu
  \right)\psi ,
\end{equation}
where
\[
\bar{\psi}\gamma^a\overleftrightarrow{D}_\mu\psi
\equiv
\bar{\psi}\gamma^a(D_\mu\psi)
-(D_\mu\bar{\psi})\gamma^a\psi .
\]
In the forward limit, the induced spin transition is chirality
preserving and proportional to the identity, so graviton exchange
produces no magic.

In summary, the zero-magic-production condition holds to good
accuracy for the interactions considered, with the notable exception
of $Z$-boson exchange in the quark sector. 

\section{Conclusions}
\label{sec:conclusions}

We have studied magic production in the forward scattering of two
distinct Standard Model fermions, taking the initial stabilizer states
to be defined in the chirality basis. In the single-mediator-exchange
approximation, magic production vanishes for gluon, photon, $W$-boson,
Higgs-boson, and graviton exchange. For $Z$-boson exchange in purely
leptonic scattering, it also vanishes at $\sin^2\theta_W=1/4$, a value
close to its phenomenological low-energy value.

At the same value of the weak mixing angle, $Z$ exchange involving
elementary quarks generally produces nonzero magic. For high-energy
forward scattering with proton external states, however, the effective
proton coupling restores zero magic in proton--proton scattering; the
corresponding condition is only approximately satisfied for neutrons.
This result does not generally extend to baryons containing strange
quarks. At low energies, the leading spin-independent nucleon
interaction in the spin basis produces no magic, whereas
spin-dependent interactions, including $Z$ exchange, can generate it.

These results point to a partial, process- and basis-dependent
connection between magic production, the fundamental interactions,
and the weak mixing angle. They motivate the question of whether the
construction of the fundamental forces encodes, in some channels, a
tendency to suppress magic production.

\vspace{2em}

\begin{acknowledgments}
We thank Yu-tin Huang for useful discussions and early collaboration.
This work is supported in part by the National Science and Technology
Council of Taiwan under Grants No.~115-2918-I-002-026,
No.~114-2112-M-002-008, and No.~115-2112-M-002-016-MY3.
\end{acknowledgments}

\bibliographystyle{apsrev4-1}
\bibliography{ZeroMagic}

\end{document}